\documentclass{article}
\usepackage{amsmath}
\usepackage{graphicx}
\usepackage{tabularx}
\usepackage{booktabs}
\usepackage{xcolor}
\usepackage{enumitem}
\usepackage{amssymb}
\usepackage{src/mlspconf}
\usepackage{url}
\usepackage{microtype}

\copyrightnotice{U.S.\ Government work not protected by U.S.\ copyright}

\copyrightnotice{979-8-3503-2411-2/25/\$31.00 {\copyright}2025 Crown}

\copyrightnotice{979-8-3503-2411-2/25/\$31.00 {\copyright}2025 European Union}

\copyrightnotice{979-8-3503-2411-2/25/\$31.00 {\copyright}2025 IEEE}

\toappear{2025 IEEE International Workshop on Machine Learning for Signal Processing, Aug.\ 31-- Sep.\ 3, 2025, Istanbul, Turkey}

\title{From Aesthetics to Human Preferences: Comparative Perspectives of Evaluating Text-to-Music Systems}
\name{%
   Huan Zhang$^{\star }$%
   \qquad Jinhua Liang$^{\star}$%
   \qquad Huy Phan$^{\dagger}$\thanks{The work does not relate to Huy Phan’s work at Meta.}
   \qquad Wenwu Wang$^{\ddagger}$
   \qquad Emmanouil Benetos$^{\star}$
}
\address{%
   $^{\star}$ Centre for Digial Music, Queen Mary University of London , United Kingdom \\%
   $^{\dagger}$ Reality Labs, Meta, France \\
   $^{\ddagger}$ Centre for Vision, Speech and Signal Processing (CVSSP), University of Surrey, United Kingdom
}

\begin{document}
\ninept

\maketitle

\begin{abstract}
Evaluating generative models remains a fundamental challenge, particularly when the goal is to reflect human preferences.  In this paper, we use music generation as a case study to investigate the gap between automatic evaluation metrics and human preferences. We conduct comparative experiments across five state-of-the-art music generation approaches, assessing both perceptual quality and distributional similarity to human-composed music. Specifically, we evaluate synthesis music from various perceptual dimensions and examine reference-based metrics such as Mauve Audio Divergence (MAD) and Kernel Audio Distance (KAD). Our findings reveal significant inconsistencies across the different metrics, highlighting the limitation of the current evaluation practice. To support further research, we release a benchmark dataset comprising samples from multiple models. This study provides a broader perspective on the alignment of human preference in generative modeling, advocating for more human-centered evaluation strategies across domains.
\end{abstract}


\section{Introduction}
\label{sec:intro}
With the rapid advances in generative models, a fundamental question arises: \textit{How well do these models truly perform, and how can we evaluate them more thoroughly and systematically?} While early efforts often relied on proxy objectives or handcrafted metrics~\cite{kong2019_acoustic,kreuk2023_audiogen,Zhang2024DExterModels}, the growing complexity of generative systems demands more human-centered and rigorous evaluation protocols.

Across fields such as language~\cite{deepseekai2025deepseekr1incentivizingreasoningcapability,qwen2025qwen25technicalreport}, vision~\cite{zhu2025dspo}, and speech~\cite{seedtts2024seedtts}, aligning model outputs with human preference has emerged as a promising direction for both training and evaluation. In the domain of audio and music, a series of models have been proposed to enhance the generation quality by incorporating human feedback, such as Tango2~\cite{Majumder2024}, BATON~\cite{liao2024baton}, DRAGON~\cite{bai2025dragon}, and SMART~\cite{jonason2025smart}. These systems typically introduce a reward model trained on human-annotated preference datasets, guiding the generator toward outputs perceived as more desirable. Preference data can be collected in various formats, including individual scores or pairwise rankings~\cite{bai2025dragon,jonason2025smart}, often reflecting subjective criteria such as music fidelity and content enjoyment. Despite these advances, understanding how well these human-centered methods generalize — especially in perceptually complex domains like music — remains an open question.

In this paper, we use music generation as a case study to provide a broader view of human-centric evaluation challenges. Specifically, we focus on the consistency and bias of different evaluation methods, benchmarking five recent text-to-music models using both human ratings and automatic reference-based scores.


This paper is structured around \textbf{three main hypotheses}, each investigated through dedicated experiments:
\begin{itemize}[leftmargin=*, itemsep=0pt, topsep=0pt]
    \item \textbf{First}, we ask whether \textit{existing human-annotated preference datasets} align with aesthetic predictors trained independently on the human aesthetic annotations.
    \item \textbf{Second}, we benchmark \textit{state-of-the-art text-to-music models} in different perceptual dimensions to understand how performance varies under different evaluation perspectives.
    \item \textbf{Third}, we examine \textit{distribution alignment} among the generated outputs by the models as well as human-composed datasets using reference-based scores like Mauve Audio Divergence (MAD)~\cite{huang2025_aligning} and Kernel Audio Distance (KAD)~\cite{chung2025_kad}.
\end{itemize}

To our knowledge, this is the first systematic study of human preference alignment in music generation. In addition to the experimental findings, we release a benchmark dataset containing generated music samples and evaluation results to prompt transparent, reproducible, and human-centered evaluation of music generative models. We hope this research will inspire further research on evaluating and optimizing generative models to better reflect human aesthetic judgments.

\begin{table*}[]
\caption{Summary of evaluation metrics for music generation systems.}
\label{tab:summary_metrics}
\centering
\begin{tabular}{@{}lccc@{}}
\toprule
\textbf{Metric}                                   & \textbf{Modality} & \textbf{Objective}           & \textbf{Human Involvement} \\ \midrule
Inception Score (ISc)        & audio      & distribution-to-distribution & $\times$   \\
Fréchet Audio Distance (FAD) & audio      & distribution-to-distribution & $\times$   \\
Per-song FAD & audio      & instance-to-distribution & $\times$   \\
FAD-$\infty$ & audio      & distribution-to-distribution & $\times$   \\
Log-spectral Distance (LSD)  & audio      & instance-to-instance         & $\times$   \\
Kullback–Leibler (KL) Divergence                  & audio             & distribution-to-distribution & $\times$                   \\
Kernel Inception Distance (KID)                   & audio             & distribution-to-distribution & $\times$                   \\
Learned Perceptual Audio Patch Similarity (LPAPS) & audio             & instance-to-instance         & $\times$                   \\
CLAP Score                   & audio-text & instance-wise                & $\times$   \\
KAD (Kernel Audio Distance)  & audio      & distribution-to-distribution & $\times$   \\
MAUVE Audio Divergence (MAD) & audio      & distribution-to-distribution & $\times$   \\
Mean Opinion Score (MOS)     & audio      & instance-wise                & $\checkmark$ \\
Meta-AudioBox-Aesthetics     & audio      & instance-wise                & $\checkmark$ \\
MusicEval-CLAP                & audio-text & instance-wise                & $\checkmark$ \\
Human Aesthetics Preference Model (DRAGON)    & audio             & instance-wise                & $\checkmark$                 \\ \bottomrule
\end{tabular}
\end{table*}

\section{Related Works}
\label{sec:related_works}
In this work, we propose a categorization of performance evaluations for music generation systems based on three dimensions: objectives, human involvement, and modality, as summarized in Table~\ref{tab:summary_metrics}. Since our focus is on aligning music generation with human preferences, this section is organized into objective and subjective evaluations, depending on whether human annotators are involved.

\subsection{Objective evaluation}\label{subsec:objective_eval}
Evaluating synthesized audio requires human annotators to rate the candidate samples according to various perceptual criteria, which is time-consuming and labour-intensive. To alleviate the annotation cost, many existing works rely on objective metrics over subjective evaluation by applying metrics without involving human listening tests~\cite{liang2024wavcraft,liu2023wavjourney,hisariya2024bridgingpaintingsmusic,lim2024hierarchical,zhang2025renderbox,yuan2024leveraging}. For example, Kong \textit{et al.} applied Inception Score (ISc) to measure the divergence among a bunch of generated audio~\cite{kong2019_acoustic}. With regard to reference-based metrics, Fréchet Audio Distance (FAD)~\cite{kreuk2023_audiogen}, Log-spectral distance (LSD) and Kullback–Leibler (KL) divergence~\cite{yang2023_diffsound} are computed between reference and synthesized audio by feeding them into pretrained deep audio networks. Gui \textit{et al.} further improved FAD by proposing per-song FAD, which calculates an individual FAD score for each song against the reference dataset, and FAD-$\infty$, which estimates an unbiased FAD score by extrapolating the FAD to an infinite sample size~\cite{gui2024adapting}. Kernel Inception Score (KID) was used to assess the quality of generated audio by measuring the distance between ground truth distributions and synthesized audio~\cite{yuan2024leveraging}. Based on the premise that the feature within deep audio networks matches human perception better, Learned Perceptual Audio Patch Similarity (LPAPS) was used to model the distance between generated and reference audio~\cite{manor2024zeroshot,liang2024audiomorphix}. To evaluate cross-modality generation performance, the contrastive audio-language model (CLAP) score was devised to reflect the alignment of generated audio and text description~\cite{liang2024audiomorphix}. KAD~\cite{chung2025_kad} was proposed based on Maximum Mean Discrepancy (MMD) to alleviate the reliance on Gaussian assumptions and sensitivity to sample size.

\subsection{Subjective human evaluation}\label{subsec:human_eval}
Human evaluation is the most straightforward to reflect the quality of synthesized audio, as well as the golden standard for music. Many works used human annotators to evaluate the generated audio and even to optimize the model for better alignment with human preference~\cite{bai2025dragon}. 

In an existing work~\cite{ragano2023_comparison} for the evaluation of text-to-speech models, neural networks are trained to predict the Mean Opinion Score (MOS). The Meta-AudioBox-Aesthetics project \cite{tjandra2025_audiobox} proposes a guideline for annotating human aesthetics scores from four axes: Production Quality (PQ), Production Complexity (PC), Content Enjoyment (CE), and Content Usefulness (CU). In particular, the annotations are sourced from 158 raters who passed through a rater qualification program that maintains alignment with a golden set. After annotating 97k audio samples (500 hours of data, with one third being music), four predictors for non-intrusive and utterance-level automatic aesthetics prediction are trained. The predictor's output is verified to be closely correlated with human perception. 

In the particular category of instrumental music, there exist assessment models with focus on music skill such as PianoJudges~\cite{Zhang2024FromPiano}, PISA assessment \cite{Parmar2021PianoAssessment} and LLaQo~\cite{Zhang2024LLaQoAssessment}. Such models' training are heavily reliant on the annotation of professional instrumentalists or pedagogues, and their domain of expertise is hardly transferable to the evaluation of general popular music. 

The work of Huang~\textit{et al.}~\cite{huang2025_aligning} quantified multiple desiderata and tested the robustness of objective evaluation metrics with regard to synthetic degradations. Along with their pairwise-annotated dataset MusicPrefs, their proposed MAD metric was computed on representations from a self-supervised audio embedding model that's demonstrated to align better with human preference.

MusicEval~\cite{liu2025musicevalgenerativemusicdataset} project collected a dataset of 2,748 music clips generated by 31 different generative methods, each of which was prompted by 384 music descriptions. Each clip was evaluated by five annotators, and their scores were averaged. Based on the data collected, a music assessment model was proposed by finetuning a CLAP model. However, as of the time of writing, neither their dataset nor model was released to the public yet.

Similarly, Bai~\textit{et al.} proposed a human aesthetics preference model for AI-generated music by collecting 1,676 human-rated clips~\cite{bai2025dragon}. They leverage the aesthetics model, together with 20 objective evaluation metrics, to optimize a music generative model, DRAGON. Despite their improved performance on various metrics, such as FAD, it is still debatable if the objective evaluation is aligned with human preference, as discussed in~\cite{chung2025_kad}. In the symbolic music domain, Nicolas \textit{et al.} applied the Meta-AudioBox-Aesthetic predictor~\cite{tjandra2025_audiobox} as a reward to improve a piano MIDI model with group relative policy
optimization (GRPO)~\cite{jonason2025smart}.

However, the approach of obtaining human annotations for training or testing data of evaluation systems suffers from inherent subjectivity. For example, the personal preferences of the annotators would lead to disagreement, making it difficult to find underlying patterns for ``high-quality music''. A study~\cite{Zhang2024HowDataset} on the teacher's rating of piano performance quality has already reflected large disagreements even under strict grading guidelines.

\section{Human vs. Human: Do \texttt{MusicPref} Annotations Agree with Aesthetics Scores?}

Can different sources of human annotation reach a common ground? In this experiment, our aim is to understand the relationship between human-annotated pairwise preferences and the independently trained aesthetics predictor, conducting a detailed comparative analysis across their perceptual dimensions, respectively. 

Each entry in \texttt{MusicPref} consists of a pairwise comparison between two audio clips generated by different models, accompanied by human judgments of which audio is preferred for musicality and fidelity. Independently, we used the Meta-AudioBox-Aesthetics model~\cite{tjandra2025_audiobox} trained on human preferences, producing scalar scores across four aspects: Content Enjoyment (CE), Content Usefulness (CU), Production Complexity (PC), and Production Quality (PQ). For each in 2515 pairs in \texttt{MusicPref}, we computed the difference in aesthetic scores between the two audio clips and test whether this difference aligns with the direction of human preference. We evaluated the agreement between human preference direction and aesthetics score deltas using two metrics:

\begin{itemize}[leftmargin=*, itemsep=0pt, topsep=0pt]
    \item \textbf{Accuracy:} The proportion of non-tie human decisions (musicality: 2049 pairs; fidelity: 1889) where the aesthetics score delta correctly predicted the preferred audio.
    \item \textbf{Spearman correlation:} Rank correlation between human preference direction (\(+1\) for A wins, \(-1\) for B wins, \(0\) for tie) and the difference in score across the dataset.
\end{itemize}

Table~\ref{tab:accuracy} reports the accuracy of each aesthetic metric in predicting human preferences. We can observe that Content Usefulness (CU) consistently yields the highest accuracy of 62.3\% for musicality, and Production Quality (PQ) predicts higher accuracy for fidelity. However, given 50\% of the random guess baseline, the results demonstrate a very poor agreement between the human annotated pairs and learnt preference scores from neural network.


\begin{table}[h]
\caption{Prediction accuracy of aesthetics score deltas against human preferences.}
\label{tab:accuracy}
\centering
\begin{tabular}{lcccc}
\toprule
\textbf{Preference Type} & \textbf{CE} & \textbf{CU} & \textbf{PC} & \textbf{PQ} \\\midrule
Musicality & 0.604 & \textbf{0.623} & 0.483 & 0.596 \\
Fidelity   & 0.572 & 0.590 & 0.432 & \textbf{0.591} \\
\bottomrule
\end{tabular}
\end{table}

Table~\ref{tab:spearman} presents the Spearman correlation between aesthetic score deltas and human preferences, supporting the low correlation trend observed in the accuracy scores. CU achieves the highest correlation for musicality (\(r = 0.258\)) and strong performance in fidelity (\(r = 0.200\)), followed closely by CE and PQ. Again, the PC performs poorly and even exhibits a weak negative correlation with fidelity.

\begin{table}[h]
\caption{Spearman correlation between aesthetic score deltas and human preferences.}
\label{tab:spearman}
\centering
\begin{tabular}{lcccc}
\toprule
\textbf{Preference Type} & \textbf{CE} & \textbf{CU} & \textbf{PC} & \textbf{PQ} \\\midrule
Musicality & 0.237 & \textbf{0.258} & -0.002 & 0.204 \\
Fidelity   & 0.177 & \textbf{0.200} & -0.087 & 0.181 \\
\bottomrule
\end{tabular}
\end{table}


\begin{table*}
\caption{Mean ± standard deviation of aesthetics scores (CE, CU, PC, PQ) for each model. \texttt{(no low)}* is the MusicCaps GT subset that removes the recordings with ``low quality'' tag.}
\label{tab:ttm_aesthetics_scores}
\centering
\begin{tabular}{lcccc}
\toprule
\textbf{Model} & \textbf{Content Enjoyment} & \textbf{Content Usefulness} & \textbf{Production Complexity} & \textbf{Production Quality} \\
\midrule
\texttt{JASCO}         & \textbf{7.33 ± 0.60} & \textbf{7.81 ± 0.42} & 5.26 ± 0.87 & \textbf{7.80 ± 0.50} \\
\texttt{Stable-Audio-Open}   & 4.55 ± 0.56 & 6.82 ± 0.43 & 2.83 ± 0.58 & 7.01 ± 0.37 \\
\texttt{MusicGen-Large}   & 6.77 ± 1.07 & 7.57 ± 0.66 & 5.03 ± 0.79 & 7.46 ± 0.66 \\
\texttt{DiffRhythm}    & 6.81 ± 1.19 & 7.18 ± 0.86 & \textbf{6.39 ± 0.93} & 7.51 ± 0.85 \\
\texttt{YUE} & 7.08 ± 0.78 & 7.60 ± 0.37 & 5.01 ± 0.93 & 7.77 ± 0.94 \\
\texttt{MusicCaps GT} & 6.21 ± 1.41 & 6.69 ± 1.32 & 5.41 ± 1.46 & 6.93 ± 1.15 \\
\texttt{MusicCaps GT (no low)}* & 6.25 ± 1.39 & 6.75 ± 1.27 & 5.44 ± 1.46 & 6.98 ± 1.12 \\
\bottomrule
\end{tabular}
\end{table*}

\section{TTM leaderboard: Who does best, and under whose judge?}
\label{sec:ttm_leaderboard}

In this section, we benchmark five recent text-to-music (TTM) generation models by evaluating their outputs on a standardized set of prompts, and reporting the four AudioBox-aesthetics metrics over synthezised music clips.

\subsection{Models}

\begin{itemize}[leftmargin=*, itemsep=0pt, topsep=0pt]
    
    \item \texttt{JASCO:} A large-scale model optimized for musical generation with strong coherence and aesthetic fluency, tuned for structured control including melody, chord and drum track~\cite{Tal2024JointGeneration}. Its output is 10 seconds. 
    
    \item \texttt{Stable-Audio-Open:} A diffusion-based model developed by Stability AI, trained on free-licensed music~\cite{Evans2024StableOpen}. The output is 47 seconds. 
    
    \item \texttt{MusicGen:} A Transformer-based model from Meta AI for TTM generation, trained on licensed music~\cite{copet2023simple}. We use the \texttt{Large} model in our experiments and their output is 20 seconds. 

    \item \texttt{YuE:} A two-stage auto-regressive system for teh generation of full-length songs~\cite{yuan2025yue}. In practice, its output is around 50 seconds depending on the length of the input lyrics. 
    
    \item \texttt{DiffRhythm:} A recent diffusion-based song generation framework that excels in inference speed~\cite{ning2025diffrhythm}. We generate outputs of 95 seconds from this model.

\end{itemize}

\subsection{Dataset}

We chose LP-MusicCaps \cite{Doh2023LP-MusicCapsCaptioning} as the prompt to generate a fixed set of music by various models, and the prompts were derived from the original tag from the MusicCaps dataset. We looked at the entire training and testing set combined, a total of 5,521 prompts. We also included the audio ground truth of the MusicCaps (human-made recording, 10s each) in the aesthetics evaluation. 


With the growing emphasis on fine-grained control in music generation, several of the TTM models evaluated in this study require additional conditioning inputs.

For models such as \texttt{DiffRhythm} and \texttt{YuE}, lyrics are required as input modality. To support this, we generated a set of 50 lyrics using \texttt{GPT-4o} spanning diverse themes and structured them according to common popular song forms that each model is capable of handling. The \texttt{JASCO} model, on the other hand, requires both drum tracks and chord progressions as additional input. We fetched a collection of 50 drum tracks from \textit{The Drum Tamer}\footnote{\url{https://www.adoptabeat.co.uk}}, covering a wide range of genres, tempi, and drum kit styles. For chord progression inputs, we generated a set of 100 symbolic chord sequences using \texttt{GPT-4o} that encompass a variety of harmonic patterns and musical styles.

\subsection{Main Results}

We report the average (\textit{mean}) and variability (\textit{standard deviation}) of the aesthetics scores along four dimensions for each model, as summarized in Table~\ref{tab:ttm_aesthetics_scores}. Notably, \texttt{JASCO} consistently achieves the highest scores across most dimensions, particularly in CU ($7.81 \pm 0.42$) and PQ ($7.80 \pm 0.50$), suggesting its strength in generating coherent and stylistically rich compositions that are perceived as both creative and high-quality. \texttt{DiffRhythm}, although a newer diffusion-based model, shows strong performance in PC ($6.39 \pm 0.93$), outperforming all baselines in this category, and remains competitive in CE and PQ. This indicates its potential in capturing fine-grained rhythmic structure and production fidelity, aided by rhythm-aware conditioning.

In contrast, \texttt{Stable-Audio-Open} underperforms in CE and PC, with relatively low scores ($4.55 \pm 0.56$ and $2.83 \pm 0.58$, respectively), although it maintains a high PQ score ($7.01 \pm 0.37$), suggesting that the outputs retain aesthetic appeal despite weaker structural or conceptual expression. \texttt{MusicGen-Large} offers a more balanced profile, trailing behind \texttt{JASCO} in CU and PQ, but outperforming \texttt{Stable-Audio-Open} across all metrics.

The ground truth reference, \texttt{MusicCaps GT}, serves as a human baseline, but is surpassed by both \texttt{JASCO} and \texttt{MusicGen-Large} in several dimensions. 
Given that a non-trivial amount of low quality recordings exists in the MusicCaps dataset (under the ``low quality'' tag or captions), we have also computed the aethetics score for the subset without the ``low quality'' tag, totaling 4384 recordings. However, the \texttt{MusicCaps GT (no low)} subset does not achieve a score that is significantly higher than the full set.

It is worth noting that some models, such as \texttt{JASCO}, \texttt{DiffRhythm}, and \texttt{Yue}, benefit from additional conditioning inputs (e.g., chord progressions, drum tracks, or lyrics), which may provide them with an advantage in generating more structured or expressive outputs. While this reflects their intended usecases, it also introduces variability in input richness, which should be considered when interpreting the comparisons.

\subsection{Is our aesthetics preference leaned towards particular kinds of music?}

\begin{figure}[h]
    \centering
    \includegraphics[width=\linewidth]{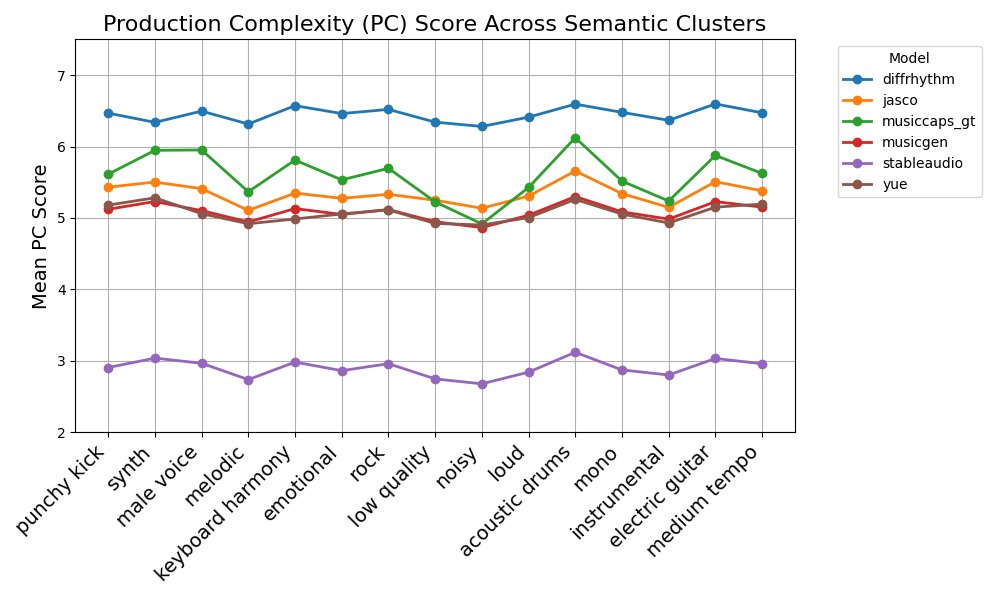}
    \caption{Production Complexity (PC) score across semantic clusters, labeled by most frequent tag. While all models follow a similar score pattern across clusters, the human reference shows greater variance.}
    \label{fig:pc_scores_by_cluster}
\end{figure}

To investigate whether the learned aesthetics metrics exhibit bias with respect to specific music types, we cluster descriptive tags (from the original tags of LP-MusicCaps) using sentence-level embeddings from a pre-trained language model (MiniLM) \footnote{\url{https://huggingface.co/sentence-transformers/all-MiniLM-L6-v2}} and KMeans clustering. Each tag is assigned to one of 15 semantic clusters, and we compute the most frequent tag within each cluster as its representative label. These labels are used on the x-axis in Figure~\ref{fig:pc_scores_by_cluster} to improve interpretability.

Among the four aesthetics dimensions, we only show the \textbf{Production Complexity (PC)} demonstrates the most prominent separation between models and cluster types, and other dimensions such as CE or CU show comparatively little inter-model or inter-cluster variability.

Figure~\ref{fig:pc_scores_by_cluster} compares PC scores across clusters for all models. Interestingly, despite architectural and training differences, most models exhibit \textit{similar scoring patterns across clusters}. This suggests that our learned aesthetics scoring function is systematically influenced by certain types of music content—clusters associated with rhythmic or electronic elements (e.g., ``punchy kick'', ``synth'') tend to receive higher PC scores, while clusters with simpler or lo-fi instrumentation (e.g., ``low quality'', ``mono'') receive lower scores.

Moreover, the ground-truth human-composed reference dataset (\texttt{musiccaps\_gt}) displays higher variance in PC scores across clusters. This could reflect the broader range of styles and production techniques in real-world music compared to the more homogeneous outputs of generative models. These findings highlight the importance of considering semantic content when interpreting evaluation scores and suggest that aesthetic predictors may encode latent content preferences shaped by the training data or annotator biases.

\section{Diversity or Drift? Dataset-Level Alignment Using Reference-Based Scores}

While the previous experiments centered on aesthetics scores and pairwise human preference alignment, this section investigates the distributional alignment among each generated music sets and human-composed set. We examine reference-based evaluation metrics on the generated dataset computed in Section~\ref{sec:ttm_leaderboard}, offering a complementary perspective to subjective assessments. Specifically, we report results using two metrics: MAD~\cite{huang2025_aligning} and KAD~\cite{chung2025_kad}. 

MAD \cite{huang2025_aligning} combines the KL divergence-based notion to measure the distributional divergence between generated and reference audio embeddings. A lower MAD score indicates a closer alignment to the real music distribution. 
KAD \cite{chung2025_kad}, on the other hand, quantifies the average kernelized distance between the generated and reference distributions in a learned perceptual space. We employ the \texttt{panns-wavegram-logmel}~\cite{Kong2020panns} as reference embedding and follows the implementation used in original repository.  

\begin{figure}[h]
    \centering
    \includegraphics[width=0.9\linewidth]{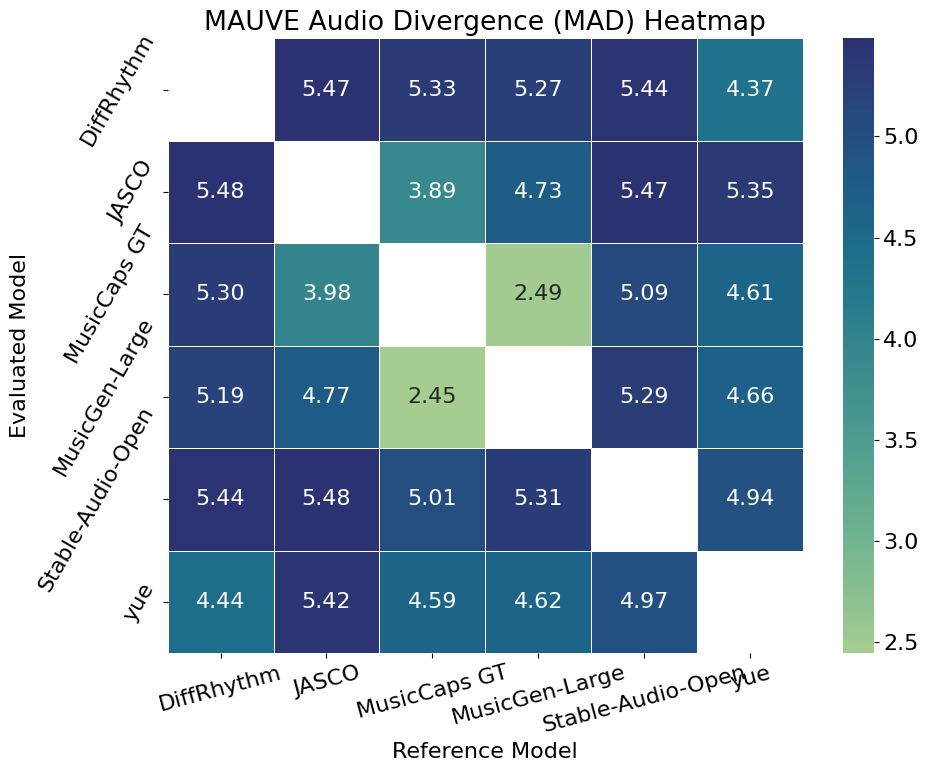}
    \caption{Asymmetric heatmap of pairwise MAD scores across the evaluated models and the ground-truth MusicCaps dataset. Lower scores indicate closer distributional alignment.}
    \label{fig:mad-heatmap}
\end{figure}

\begin{figure}[h]
    \centering
    \includegraphics[width=0.9\linewidth]{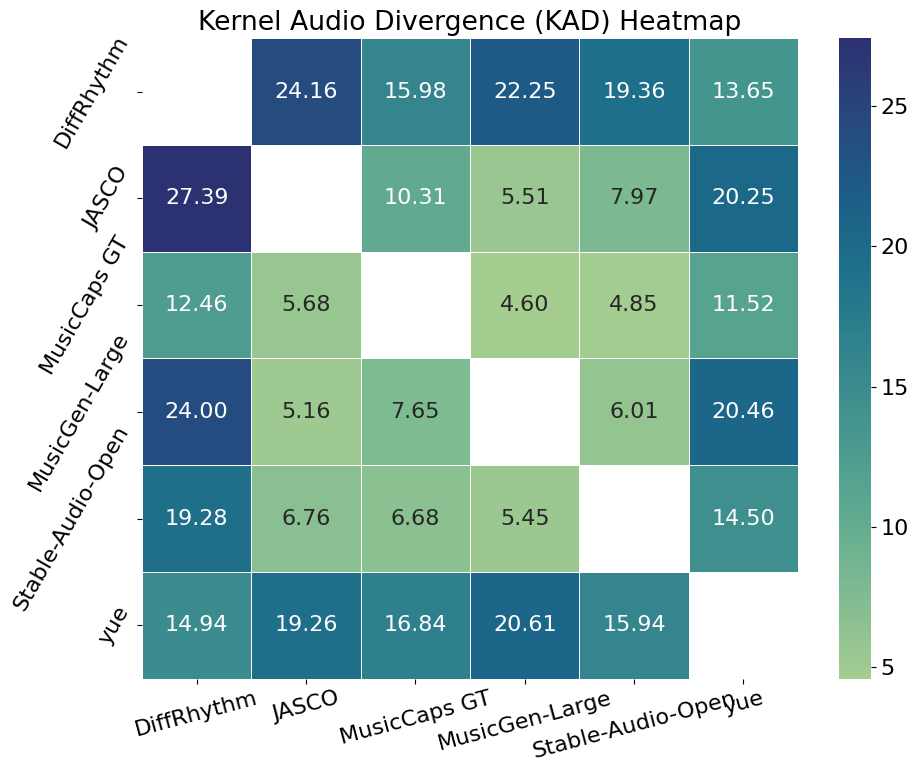}
    \caption{Pairwise KAD scores between the evaluated models and the MusicCaps ground-truth dataset. Lower scores indicate closer proximity in the learned perceptual space.}
    \label{fig:kad-heatmap}
\end{figure}

Figure~\ref{fig:mad-heatmap} visualizes the asymmetric pairwise MAUVE Audio Divergence (MAD) scores across the models' outputs. Notably, \texttt{MusicGen-Large} exhibits the lowest MAD score with respect to \texttt{MusicCaps GT} (2.45), followed by \texttt{JASCO} (3.89) when evaluated against the other models. These results suggest that \texttt{MusicGen-Large} is best aligned with the reference distribution, while models such as \texttt{DiffRhythm}, \texttt{Stable-Audio-Open}, and \texttt{YuE} exhibit consistently higher MAD scores, indicating larger distributional shifts.

The Kernel Audio Distance (KAD) scores in Figure~\ref{fig:kad-heatmap} further corroborate these findings. Again, \texttt{MusicGen-Large} and \texttt{JASCO} achieve the lowest distances relative to the \texttt{MusicCaps GT} reference (7.65 and 5.51, respectively), underscoring their strong alignment with perceptual aspects of real music. In contrast, \texttt{DiffRhythm}, \texttt{Stable-Audio-Open}, and especially \texttt{YuE} show substantially higher distances from all other sets, suggesting that their output distributions deviate more significantly. Comparatively, the \texttt{DiffRhythm} and \texttt{YuE} are relatively close in terms of both MAD and KAD values, which is explainable given that both models have a focus on song generation.

\section{Conclusion and Future Work}
\label{sec:conclusion}

In this work, we presented a cross-referenced discussion of text-to-music generation evaluations. Our results reveal substantial inconsistencies between different evaluation perspectives, highlighting the challenges of fully capturing human judgment through automated proxies. By benchmarking five recent models and releasing the generated dataset, we aim to support more transparent, reproducible, and human-centric evaluation practices in music generation research. Future work includes exploring richer annotation schemes and developing evaluation methods that better reflect detailed music preferences that varies across culture background, music training, listening habits, etc.



\bibliographystyle{src/IEEEbib}
\bibliography{src/refs, ref}

\end{document}